\documentclass{PoS}

\usepackage{multirow}

\newcommand{\C}[1]{{\cal C}_{#1}}

\title{$B \to K^*(\to K\pi) \ell^+\ell^-$ theory and the global picture: What's next?}

\ShortTitle{$B \to K^*(\to K\pi)\ell^+\ell^-$ theory and the global picture: What's next?}

\author{B. Capdevila\\
Universitat Aut\`onoma de Barcelona and IFAE,\\ 08193 Bellaterra, Spain}
\author{S. Descotes-Genon\\
Laboratoire de Physique Th\'eorique  (UMR 8627), CNRS, Univ. Paris-Sud,\\ Universit\'e Paris-Saclay, 91405 Orsay Cedex, France}
\author{L. Hofer\\
Universitat de Barcelona (UB), Mart\'i Franqu\`es 1,\\ 08028 Barcelona, Spain}
\author{\speaker{J. Matias}\\
Universitat Aut\`onoma de Barcelona and IFAE,\\ 08193 Bellaterra, Spain}
\author{J. Virto\\
Albert Einstein Center for Fundamental Physics, Institute for Theoretical Physics,\\
University of Bern, CH-3012 Bern, Switzerland.}

\abstract{The present status of the LHC anomalies found in exclusive semileptonic $b\to s\ell^+\ell^-$ decays
is discussed with special emphasis on the exclusive 4-body angular distribution $B \to K^*(\to K\pi)\ell^+\ell^-$. The treatment of hadronic uncertainties in this mode is briefly reviewed, and some of the analyses in the literature  are critically reassessed. The global picture provided  by the global fit points to a coherent pattern of deviations with a significance substantially above 4$\sigma$ for different New Physics scenarios. Finally, we propose as the next step in the field to  focus on the study of optimized observables that compare electron and muon modes, sensitive to lepton-flavour universality violations and free from hadronic uncertainties (including charm) in the SM, the so called $Q_i$ observables.}

\FullConference{Fourth Annual Large Hadron Collider Physics\\
		13-18 June 2016\\
		Lund, Sweden}

\begin{document}
Flavour Physics and particularly rare B decays is slowly evolving from a Precision Flavour Physics Era to a New Physics Search Era. The lack of any solid evidence of New Physics (NP) in direct searches  has put Flavour Physics at the forefront of the main NP searches. Several evidences have slowly  piled up in a coherent pattern of deviations in the last three years. Not all of them 
 have been tested with the same degree of accuracy using the latest data and different experiments. For instance, the observable $R_K=Br(B \to K\mu^+\mu^-)/Br(B \to K e^+e^-)$ that exhibits a deviation of 2.6$\sigma$ w.r.t. SM prediction  and points to a universal lepton flavour violation (LFUV) in $b \to s$ transitions, has been measured by LHCb\cite{rk} only 
 using data corresponding to an integrated luminosity of  1 fb$^{-1}$ (an independent measurement by Babar is also available \cite{babarrk}). More statistics is required to fully confirm it\footnote{Another interesting  LFUV signal has been observed in the ratio ${\rm Br}(B \to D^{(*)} \tau\nu_\tau)/{\rm Br}(B \to D^{(*)} \ell\nu_\ell$)  the so called $R(D)$ and $R(D^*)$ tension \cite{belle,lhcb}. Even if they belong to a different quark transition it is reasonable under general arguments to expect that a violation of lepton flavour universality, if real, would manifest in different type of transitions. It remains to be seen if all these hints can be accommodated into a unified model of LFUV for all three generations.}. On the contrary, the observable $P_5^\prime$ \cite{p5p} that belongs to a set of optimized observables for $B\to K^*\mu^+\mu^-$~\cite{opt1,opt2,opt3} has been measured by LHCb~\cite{2013lhcb} with 1 fb$^{-1}$ finding a 3.7$\sigma$ deviation in the long bin [4.3,8.68], later confirmed again by LHCb~\cite{2015lhcb} with the 3 fb$^{-1}$ dataset with a 3$\sigma$ deviation in each of two adjacent bins in the dilepton invariant mass [4,6] and [6,8] and recently Belle~\cite{belle2016} in a long bin [4,8] found a result nicely consistent with LHCb (see Fig.1). Also the $Br(B_s \to \phi \mu^+\mu^-)$ has been measured by LHCb \cite{lhcbphi} with 3 fb$^{-1}$ dataset exhibiting $\gtrsim 2\sigma$ w.r.t. the SM~\cite{bsz} (albeit highly sensitive to hadronic uncertainties) in different large and low recoil bins.

The crucial point that has attracted a lot of  attention from the community is that all the observed tensions ($P_5^\prime$, $R_K$, $Br(B_s \to \phi\mu^+\mu^-)$ and several low-recoil bins of $B \to P\ell^+\ell^-$ and $B \to V\ell^+\ell^-$ decays) come from observables that share the same effective couplings, the Wilson coefficients ${\cal C}_{7,9,10}^{(\prime)}$
associated with the local four-fermion operators in the effective Hamiltonian:
\begin{eqnarray}
\!\!\!  &&\mathcal{O}_9^{(\prime)}=\frac{\alpha}{4\pi}[\bar{s}\gamma^\mu P_{L(R)}b]
   [\bar{\mu}\gamma_\mu\mu],\quad
     \mathcal{O}_{10}^{(\prime)}=\frac{\alpha}{4\pi}[\bar{s}\gamma^\mu P_{L(R)}b]
   [\bar{\mu}\gamma_\mu\gamma_5\mu],\quad
  \mathcal{O}_7^{(\prime)}=\frac{\alpha}{4\pi}m_b[\bar{s}\sigma_{\mu\nu}P_{R(L)}b]F^{\mu\nu},\nonumber\\
\!\!\!  &&\mathcal{C}_9^{\rm SM}(\mu_b)=4.07,\qquad\qquad\qquad \mathcal{C}_{10}^{\rm SM}(\mu_b)=-4.31,\qquad\qquad\qquad\qquad  \mathcal{C}_7^{\rm SM}(\mu_b)=-0.29,
\end{eqnarray}
where $P_{L,R}=(1 \mp \gamma_5)/2$, $m_b$ denotes the $b$ quark mass, $\mu_b=4.8$ GeV, and
primed operators have vanishing or negligible Wilson coefficients in the SM.

\begin{figure} \label{Bellef}\begin{center}
\includegraphics[width=6.5cm,height=4.5cm]{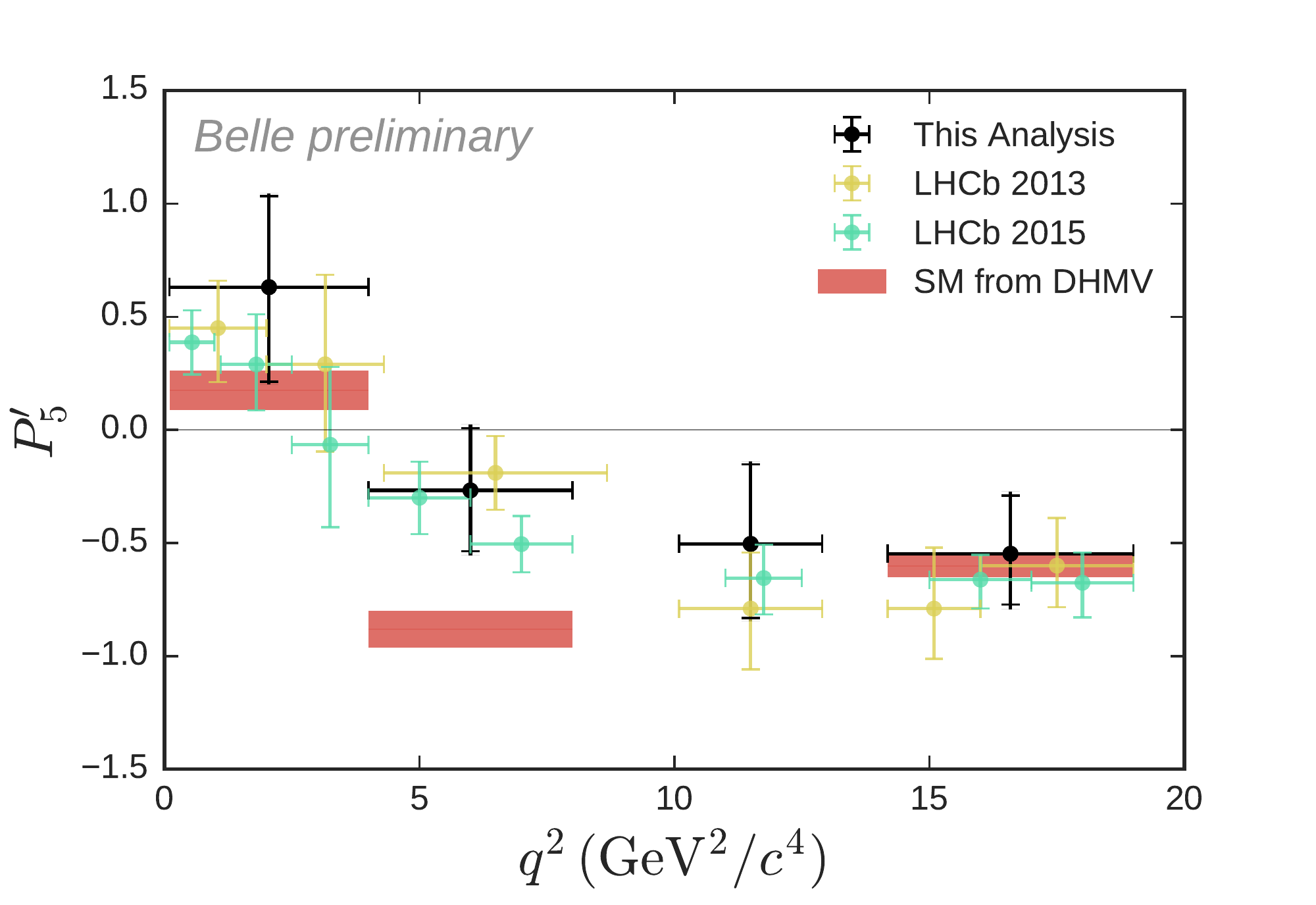}\,
\includegraphics[width=6.5cm,height=4.5cm]{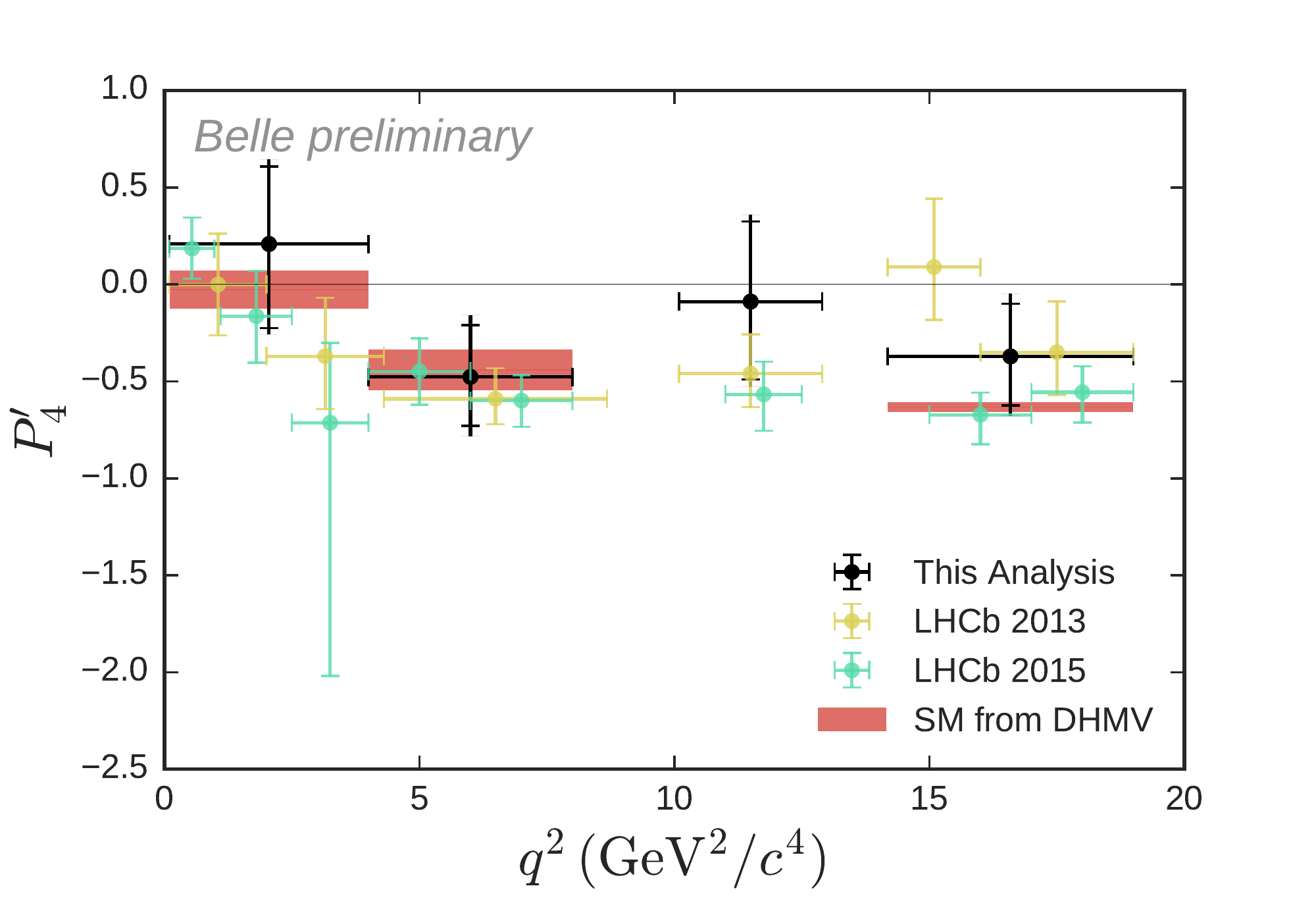}
\caption{LHCb and Belle data on $P_5^\prime$ and $P_4^\prime$ compared to DHMV predictions\cite{global}.}
\end{center}\end{figure}

The natural question to ask then is whether there exists a consistent/common NP contribution (for instance a $Z^\prime$ or leptoquarks) to these Wilson coefficients that can account and alleviate all the tensions at once. In order to answer this question in a model-independent way a global fit to all relevant processes provides a systematic way  to identify (if exists) such a common pattern. The result of the global fit is briefly discussed in Section 1.
One of the key ingredients in this fit is the angular distribution of $B \to K^*(\to K\pi)\ell^+\ell^-$, a detailed description of the  structure of this decay and the treatment of hadronic uncertainties  will be reviewed in Section 2. Finally, we discuss the next step and prospects in this research field for the near future in Section 3.

\section{Global fits}

The rare $b \to s \ell \ell$ decays allow one to determine with precision the Wilson coefficients where 
 potentially NP is encoded.
The importance of this kind of analysis to get a global picture has been clear for a long time\cite{bulk}. The first analysis exploiting the LHCb 2013 data~\cite{anomaly} pointed to a large negative contribution to the Wilson coefficient ${\cal C}_9$. This picture was confirmed later on by other groups~\cite{Altmannshofer:2013foa,Beaujean:2013soa} using different observables, Form Factors (FF), statistical and theoretical approaches.


The updated analysis of Ref.\cite{global} improves on Ref.\cite{anomaly} in many respects: it includes the latest theory and experimental updates of all relevant decays, it uses refined and accurate techniques to estimate hadronic uncertainties from factorizable power corrections to non-perturbative charm-loop non-factorizable contributions, 
and includes theory and experimental correlations.
{\tiny
\begin{table}[b]
 \begin{center}
\begin{tabular}{@{}crccc@{}} 
\hline
Coefficient ${\cal C}_i^{NP}={\cal C}_i-{\cal C}_i^{SM}$ & Best fit & 1$\sigma$ & 3$\sigma$ & Pull$_{\rm SM}$ \\ 
 \hline
   $\C7^{\rm NP}$ & $ -0.02 $ & $ [-0.04,-0.00] $ & $ [-0.07,0.03] $ &  1.2   \hspace{5mm}  \\[2mm] 
  $\C9^{\rm NP}$ & $ -1.11 $ & $ [-1.31,-0.90] $ & $ [-1.67,-0.46] $ &   4.9  \hspace{5mm}
    \\[2mm] 
 $\C{10}^{\rm NP}$ & $ 0.61 $ & $ [0.40,0.84] $ & $ [-0.01,1.34] $ &  3.0 \hspace{5mm}  \\[2mm] 
    $\C{7'}^{\rm NP}$ & $ 0.02 $ & $ [-0.00,0.04] $ & $ [-0.05,0.09] $ &  1.0  \hspace{5mm}  \\[2mm]   
       $\C{9'}^{\rm NP}$ & $ 0.15 $ & $ [-0.09,0.38] $ & $ [-0.56,0.85] $ &  0.6 \hspace{5mm}  \\[2mm] 
  $\C{10'}^{\rm NP}$ & $ -0.09 $ & $ [-0.26,0.08] $ & $ [-0.60,0.42] $ &  0.5 
 \hspace{5mm}  \\[2mm] 
  $\C9^{\rm NP}=-\C{10}^{\rm NP}$ & $ -0.65 $ & $ [-0.80,-0.50] $ & $ [-1.13,-0.21] $ &   4.6 \hspace{5mm}
    \\[2mm] 
   $\C9^{\rm NP}=-\C{9'}^{\rm NP}$ & $ -1.07 $ & $ [-1.25,-0.86] $ & $ [-1.60,-0.42] $ &  4.9 \hspace{5mm}
  \\[2mm] 
  $\C9^{\rm NP}=-\C{10}^{\rm NP}=-\C{9'}^{\rm NP}=-\C{10'}^{\rm NP}$   & $ -0.66 $ & $ [-0.84,-0.50]  $ & $ [-1.25,-0.20] $ &  4.5 
 \hspace{5mm}  \\[2mm] 
  \hline 
\end{tabular}
\end{center}
\caption{Results of one-parameter fits for the Wilson coefficients $\C{i}$ considering all data. From Ref.~\cite{global}.
\label{tab:fitres}
}
 \end{table}}
 
\begin{figure}[b]
\includegraphics[width=7.5cm,height=3.6cm]{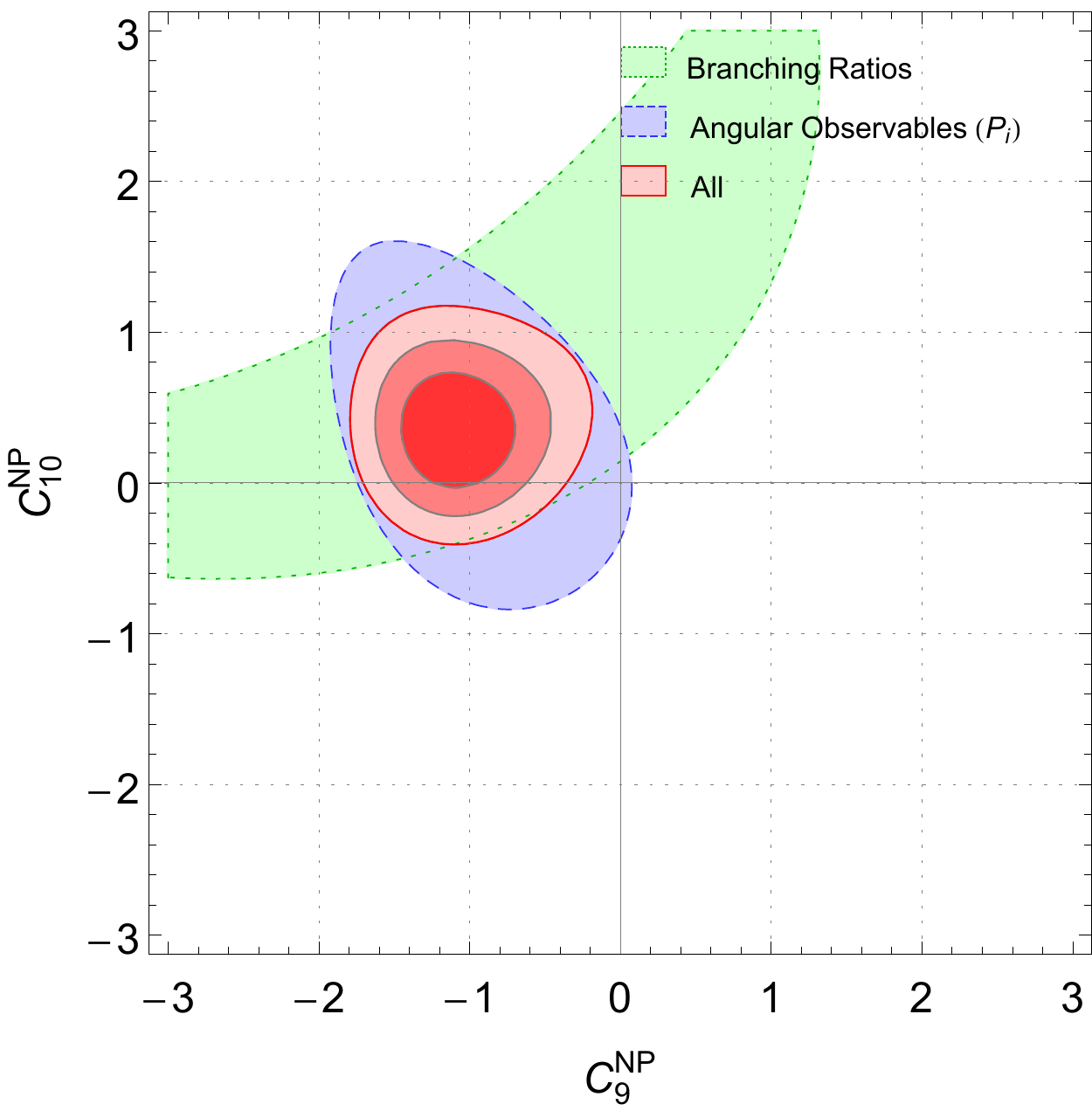}
\includegraphics[width=7.5cm,height=3.6cm]{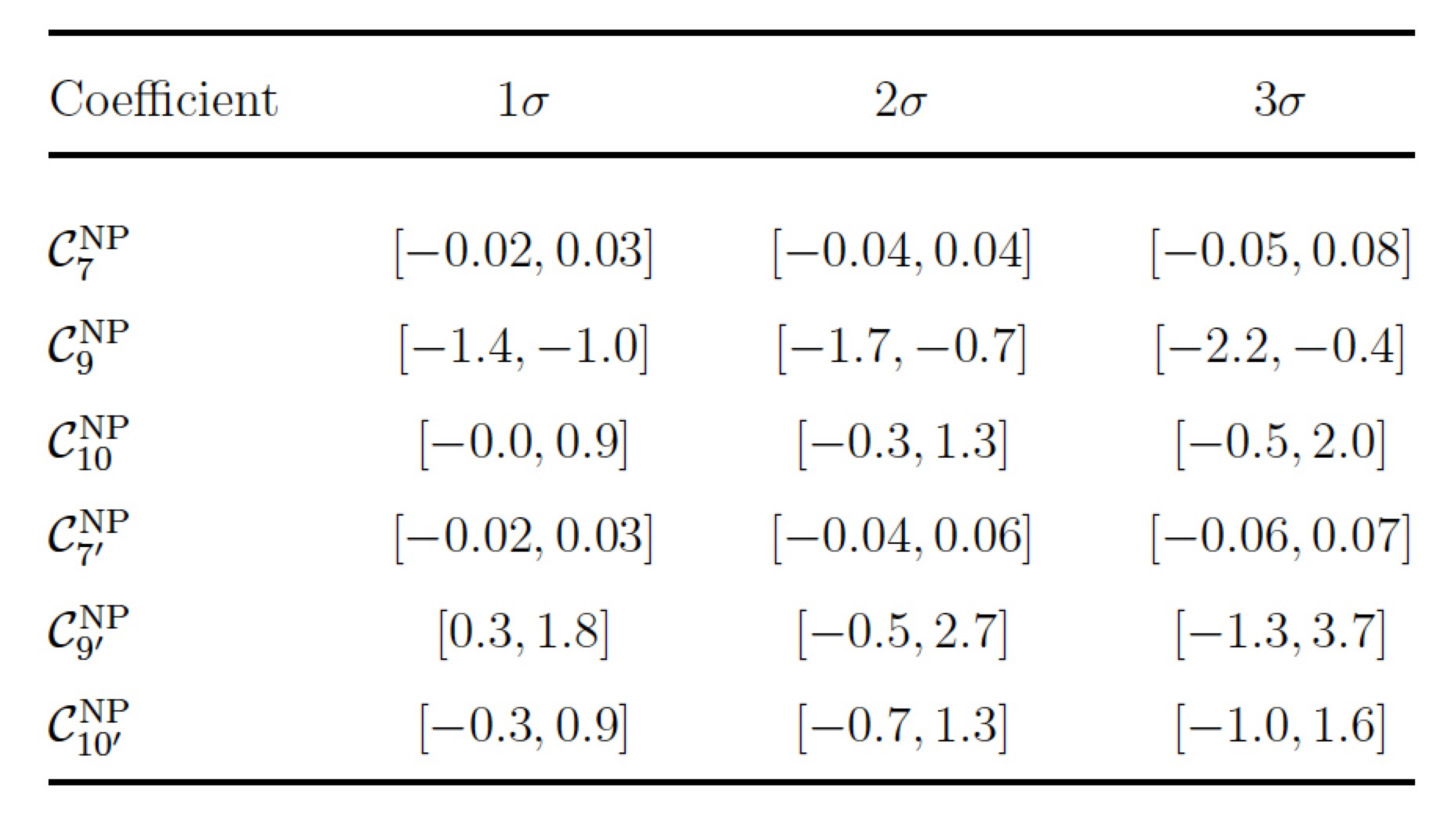}
\caption{(Left) Example of 2D fit from Ref.\cite{global} illustrating the weight of angular observables versus branching ratios. (Right) Ranges for each Wilson coefficient projected from the 6D fit\cite{global}.}\label{2dfit}
\end{figure}

\begin{figure}
\includegraphics[width=7cm,height=3.7cm]{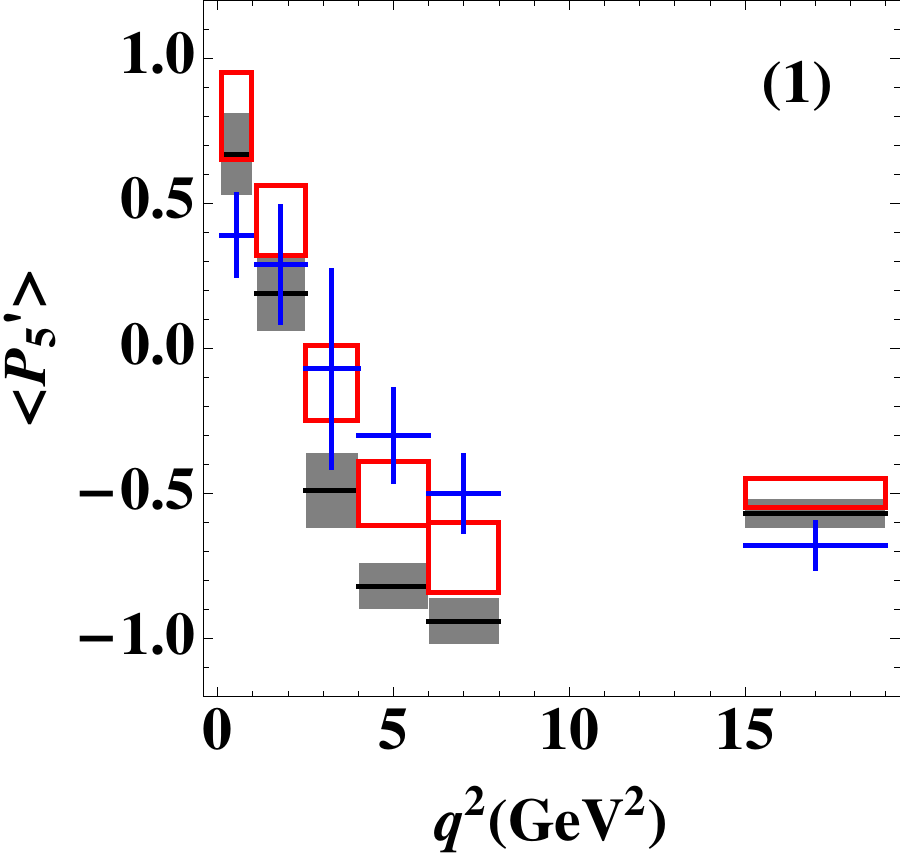}\,\,
\includegraphics[width=7cm,height=3.7cm]{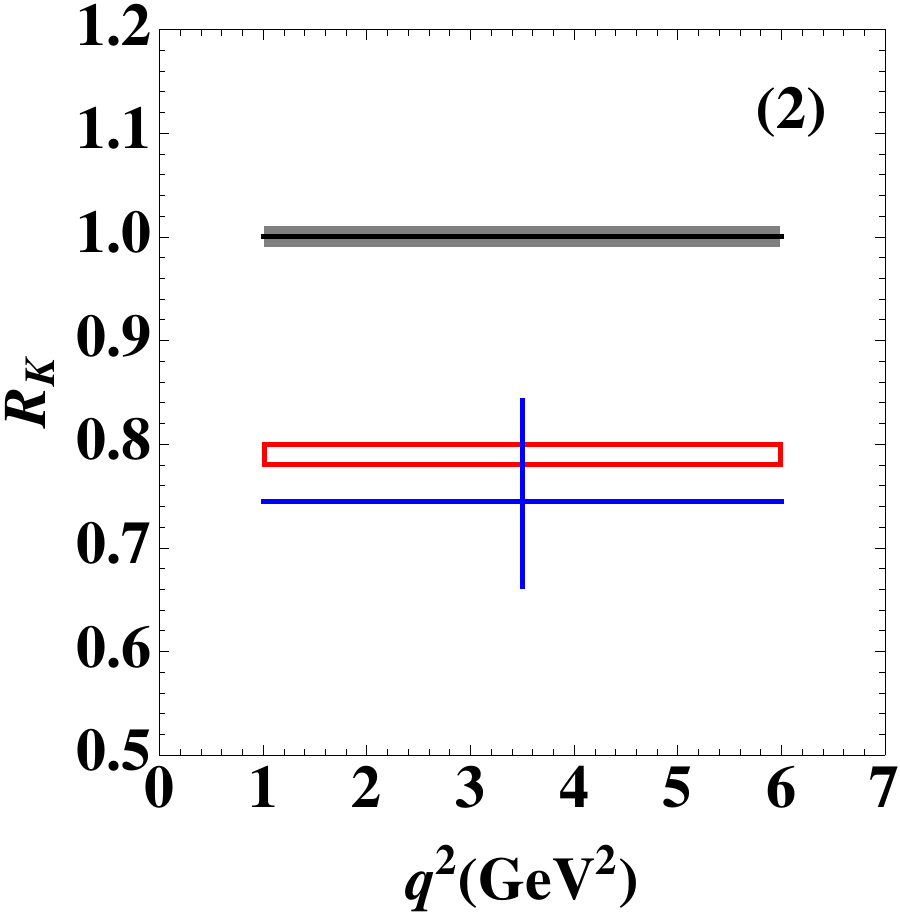}\\

\includegraphics[width=6.7cm,height=3.3cm]{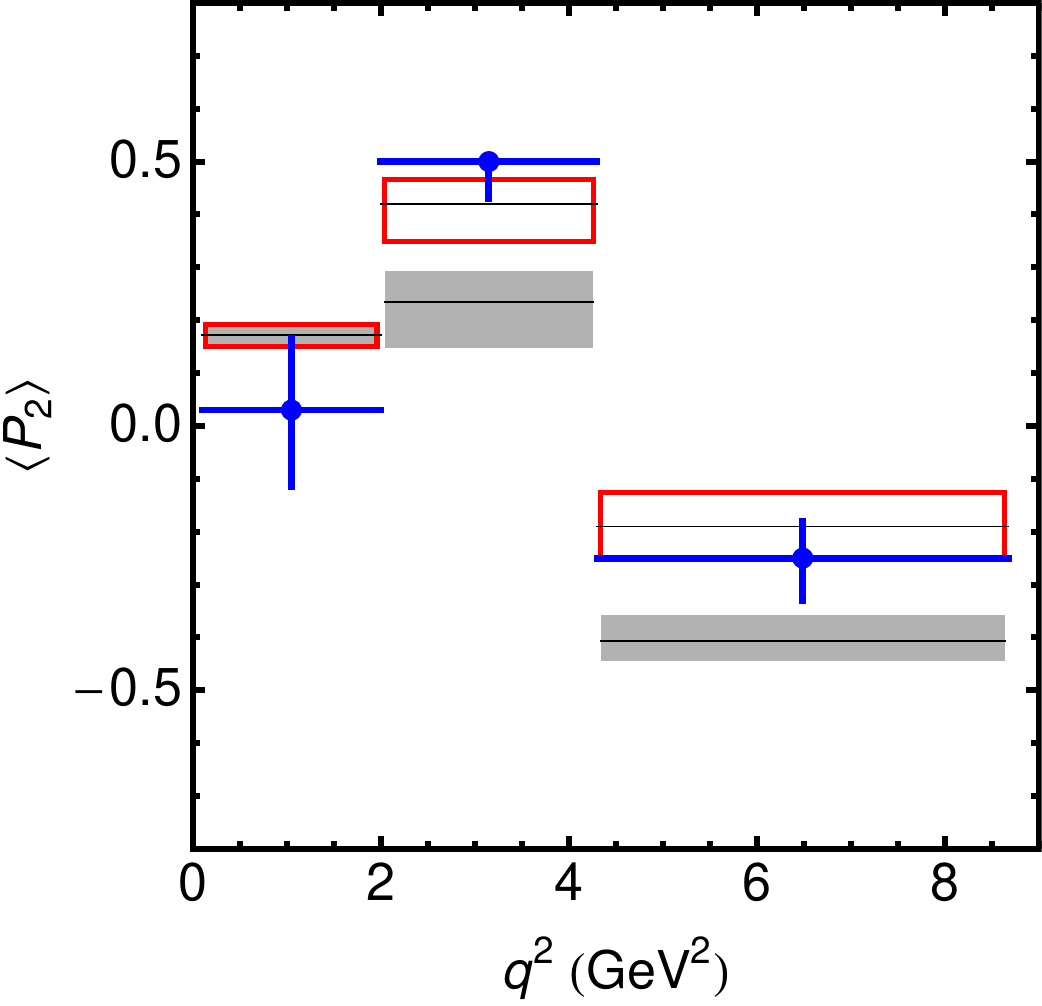}\,\hspace*{0.6cm}
\includegraphics[width=7cm,height=3.3cm]{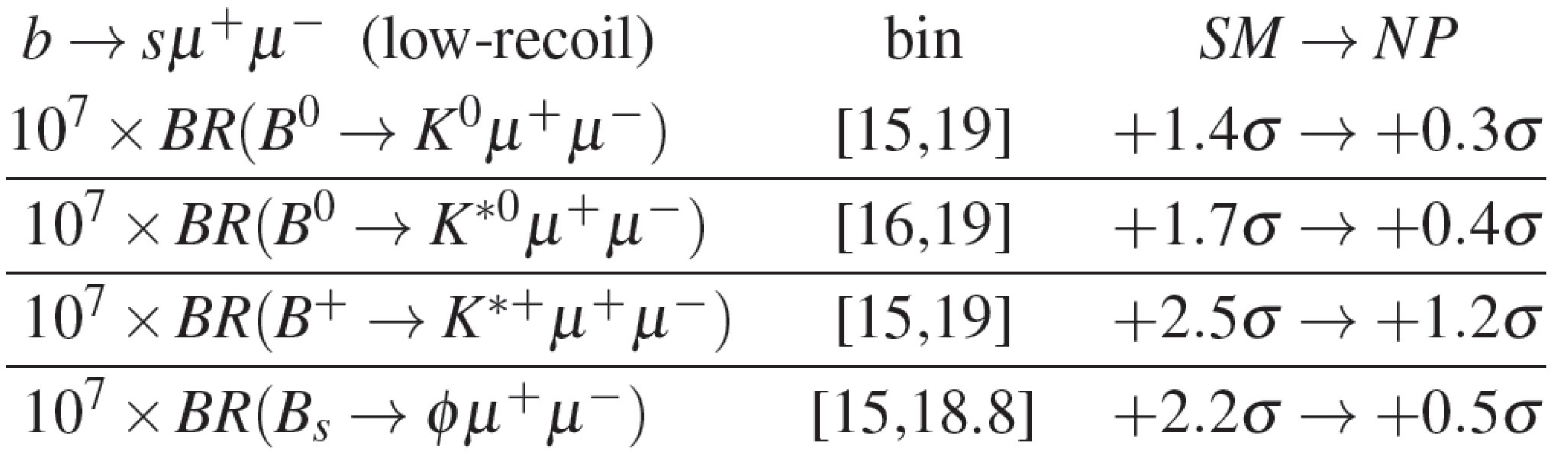}
\caption{The grey box is the SM prediction, the red box is the NP prediction for ${\cal C}_9^{NP\mu}=-1.1$ (for (1) and (2) plots) and blue crosses is data. (Left-top) The $P_5^\prime$ discrepancy is nicely alleviated. (Right-top) $R_K$ gets a good  agreement at the NP point. (Bottom-left) Old 2013 LHCb data (1fb$^{-1}$) for $P_2$  with NP point ${\cal C}_9^{NP\mu}=-1.5$ from \cite{anomaly}. (Bottom-right) Several low recoil bins see reduced their tensions if ${\cal C}_9^{NP\mu}=-1.1$ is taken.} \label{impact}
\end{figure}

Our reference fit \cite{global} includes the branching ratios and angular optimized observables for $B \to K^*\mu\mu$ and $B_s \to \phi\mu\mu$, the branching ratios of the charged and neutral modes for $B\to K\mu\mu$, the branching ratio of $B\to X_s\mu\mu$, $B_s \to\mu^+\mu^-$ and $B\to X_s\gamma$, isospin asymmetry $A_I$ as well as the time-dependent asymmetry $S_{K*\gamma}$ of $B \to K^*\gamma$. Large and low-recoil data is included for exclusive decays. Our analysis involves nearly a hundred observables in total. The SM-pull \footnote{This quantifies by how many standard deviations the best fit point is preferred over the SM point ($C_i^{NP}=0$) in a particular  scenario where a subset of Wilson coefficients is allowed to vary freely.} of our frequentist $\chi^2$ analysis indicates (as in 2013) that a large negative NP contribution to the Wilson coefficient ${\cal C}_9^{\mu}$ of order $25\%$ w.r.t. the SM value (see Table\ref{tab:fitres}) is preferred with a significance of 4.9$\sigma$ (if all data is included), 4.5$\sigma$ (without electronic mode), 3.8$\sigma$ (if only $b \to s\mu^+\mu^-$ is taken excluding [6,8] and without electronic mode, increasing to 4.3$\sigma$ if electronic mode is added). One example of the 2D fit is shown in Fig.\ref{2dfit}, exhibiting the relative weight of branching ratios versus angular observables as well as a table with the result of the 6D fit where all 6 Wilson coefficients are allowed to vary. In the table  the ranges for each ${\cal C}_i^{NP}$ are given as projections of the 6D region. The SM-pull of the 6D fit is 3.6$\sigma$.
Similar results and conclusions are obtained by Ref.\cite{straub} (see also \cite{Hurth})
 using different observables, 
 theoretical approaches and light-cone sum rule (LCSR) computations for the FF. 
Finally, it is a very illustrative exercise to show the impact of this particular NP point ${\cal C}_9^{NP\mu}=-1.1$ on some of the different anomalies and tensions (see Fig.\ref{impact}).

\medskip

\section{$B \to K^*(\to K\pi)\mu^+\mu^-$ anatomy and uncertainties}

\begin{figure}[b]
\begin{center}
   \includegraphics[width=0.25\linewidth]{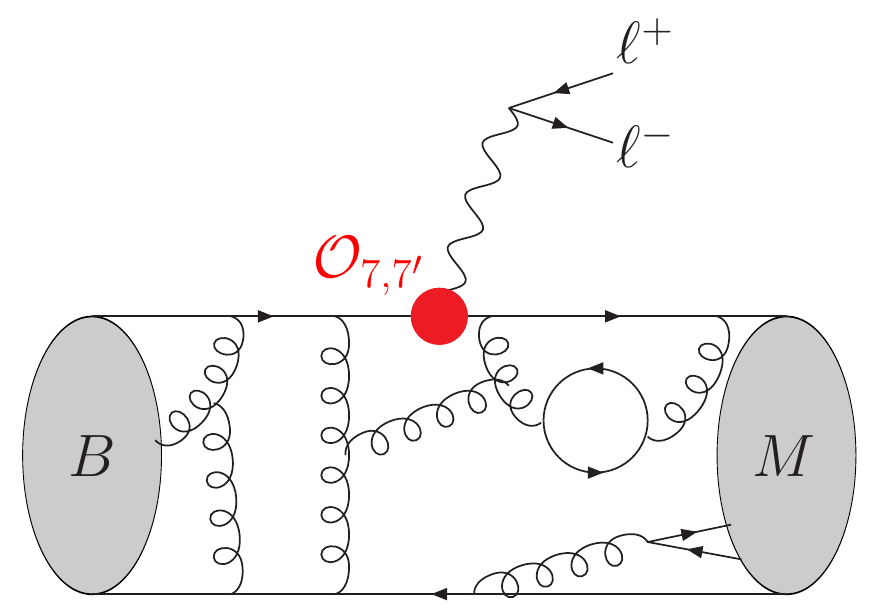}\qquad
   \includegraphics[width=0.25\linewidth]{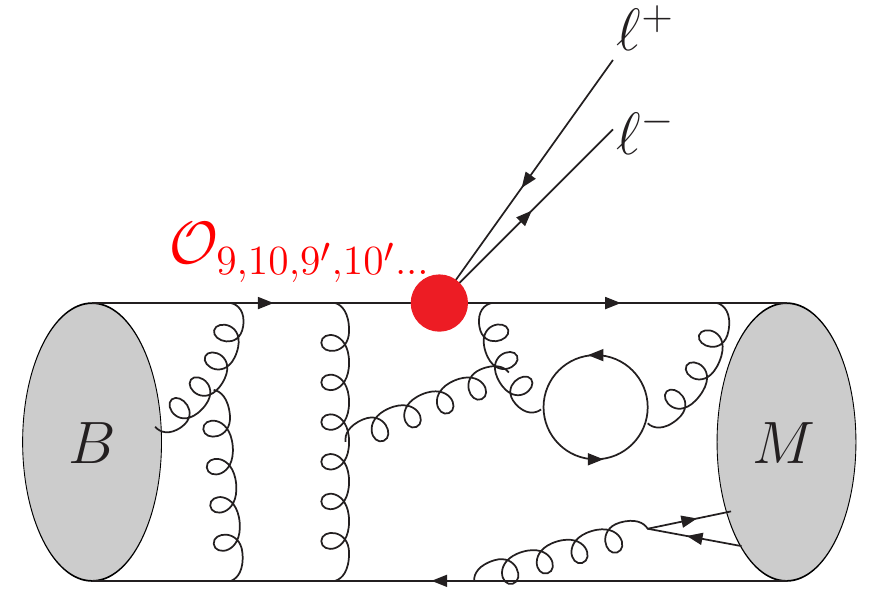} \qquad
   \includegraphics[width=0.25\linewidth]{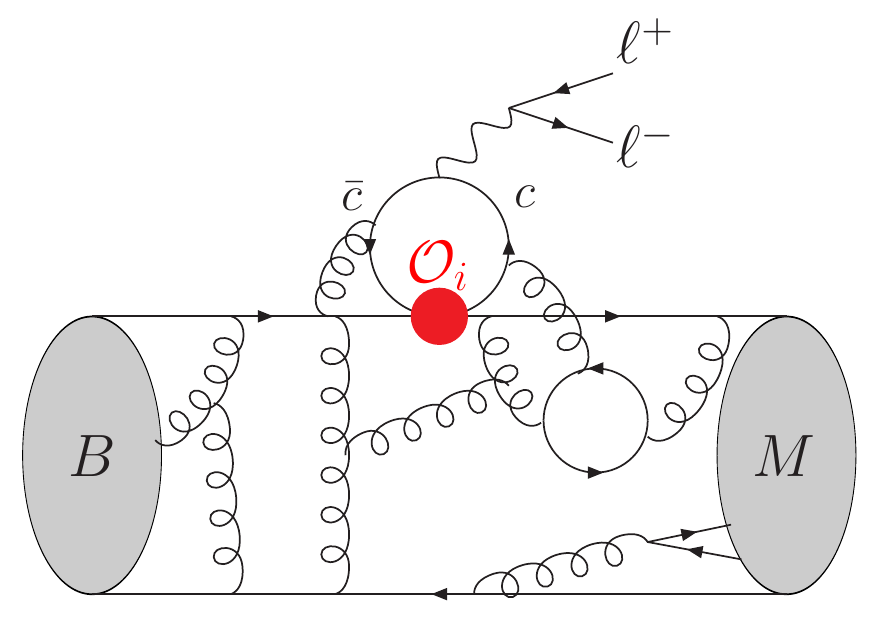}
\end{center}  
\caption{Illustration of factorisable (first two diagrams) and non-factorisable (third diagram) 
QCD corrections to exclusive $B\to M\ell^+\ell^-$ matrix elements. \label{fig:QCDcorr}}
\end{figure}

One of the key ingredients of the global fit is the angular distribution $B\to K^*(\to K\pi)\mu^+\mu^-$. The amplitude structure of this decay includes tree-level diagrams with insertions of the operators ${\cal O}_{7,9,10}$ (generated at one-loop in the SM) that contribute to the vector and axial part of the amplitude and one-loop diagrams with an insertion of the charged current operator ${\cal O}_2=[\bar s \gamma_{\mu} c_L][\bar c \gamma^\mu b_L]$ (generated at tree level in the SM) that induce a contribution to the vector part as well:
$$ { \cal A } \propto \left( {\cal A}_V^\mu + {\cal H}_V^\mu \right) \bar\ell \gamma_\mu \ell + {\cal A}_A^\mu \bar \ell \gamma_\mu\gamma_5 \ell $$
with ${\cal A}_V^\mu= C_7 \frac{2 i m_b}{q^2} q_\rho \langle \bar K^* |\bar s \sigma^{\rho\mu} b_R|\bar B\rangle + C_9  \langle \bar K^* |\bar s \gamma^{\mu} b_L|\bar B\rangle$, ${\cal A}_A^\mu=C_{10}  \langle \bar K^* |\bar s \gamma^{\mu} b_L|\bar B\rangle$ and the term ${\cal H}^\mu_V \propto i \int d^4x e^{i q . x} \langle \bar K^* |T[\bar c\gamma^\mu c] {\cal H}_c |\bar B \rangle$ where ${\cal H}_c$ involves 4-quark operators with two charm fields, i.e, ${\cal O}_2$ (see Fig.\ref{fig:QCDcorr}).
Consequently the amplitude contains local contributions yielding FF computed using non-perturbative methods like LCSR or lattice simulations, and non-local ones involving $c\bar c$ loops propagating before annihilating into a virtual photon that are estimated with different tools depending on the energy of the dilepton pair. At low-$q^2$ we work within the framework that we called  Improved QCDF  where four type of corrections are included: i) factorizable $\alpha_s$ contributions computed in QCDF \cite{thorsten, thorsten2} ii) factorizable power-corrections \cite{onthepower}  iii) non-factorizable $\alpha_s$ corrections in QCDF \cite{thorsten, thorsten2} and iv) non-factorizable power corrections  computed within LCSR with single soft gluon  contribution \cite{KMPW}. 
Factorisable corrections come  from the computation of FF within QCDF (useful to implement correlations among the FF) whereas non-factorisable arise at the level of the amplitudes themselves (and include $c\bar c$ loops among others).
These hadronic uncertainties, their treatment and a discussion of some results in the literature are commented in the following subsections. We  use for our predictions the LCSR computation of FF based on B-meson distribution amplitudes from KMPW \cite{KMPW}. A different choice is the determination of FF based on light-meson distribution amplitudes called BSZ \cite{bsz}. KMPW predictions  include BSZ  given  their very conservative error estimate an order of magnitude larger.

\subsection{Hadronic uncertainties for $B \to K^*\mu\mu$: factorizable power corrections}
It was argued some time ago \cite{jaeger} that this type of corrections might be huge and behind the anomalies observed. This issue was clarified in \cite{onthepower} and it will be explained in full detail in  \cite{newpaper}.   
 \begin{table}
 \begin{tabular}{@{}l|r|rr|@{}}
\hline\hline 
ONLY factorizable p.c error of { $\langle P_5^\prime\rangle_{[4,6]}$}  &   scheme-1 & scheme-2  \\ 
  &   {DHMV'14}\cite{onthepower} \& {BFS'05} \cite{thorsten2}&  { JC'14} \cite{jaeger2}\\ 
\hline 
OPTION i) error size of $a_F,b_F,c_F \sim 10\% \times F$,   & &  \\ {\bf only
 correlations} from large-recoil symmetries.
  & $ \pm 0.05$ & $ \pm  0.15$   \\ \hline 
OPTION ii) errors of $a_F,b_F,c_F$ including & & \\
 {\bf all correlations}  from BSZ 
 & $ \pm 0.03$ & $ \pm 0.03$   \\ \hline
 \end{tabular} \caption{Impact of different choices in the computation of factorizable power corrections for the uncertainty of $\langle P_5^\prime\rangle_{[4,6]}$. Only  factorizable p.c error of 
 $\langle P_5^\prime\rangle_{[4,6]}$ 
 is displayed. BSZ is used in this illustrative example. JC'12 and JC'14 used option i) and scheme 2 inflating the error artificially.} \label{options}
 \end{table}

 Here we will simply illustrate the main point using a simple example, but the details can be found in \cite{onthepower,newpaper}.
In QCDF, a full FF  can be decomposed in three terms:
$$F^{full}(q^2)=F^{soft}(\xi_\perp,\xi_\|)+\Delta F^{\alpha_s}(q^2)+\Delta F^{\Lambda}(q^2)$$
 a) $ F^{soft}$ is the large-recoil limit at LO in $\alpha_s$ and $1/m_b$-expansion (soft-form-factor contribution) that has embedded the leading correlations among the FF based on large-recoil symmetries \cite{charles} b) $\Delta F^{\alpha_s}(q^2)$ is the known $\alpha_s$ contribution contained in the definition of the QCD FF for heavy-to-light transitions with insertions of operators $O_7$ and $O_{1-6} $\cite{thorsten,thorsten2} and c) $\Delta F^{\Lambda}(q^2)=a_F+b_F q^2/m_B^2+c_F q^4/m_B^4$ correspond to the power corrections (first introduced by \cite{jaeger}). In our approach the central values for these coefficients $a_F, b_F,..$ can be obtained (see \cite{onthepower,newpaper}) from a fit to  specific LCSR computations  of the full FF. The results found in \cite{onthepower,newpaper} using KMPW FF were typically below $10\%$ as expected by dimensional arguments. The keypoint here is the {\it correct treatment of the uncertainty assigned  to these coefficients}  \cite{onthepower, newpaper}. In short, there are two options to assign an error to the coefficients $a_F,...$  i) a (relatively) model independent approach where the errors are taken to be uncorrelated and of order $10\% \times F$ (this is the approach followed by \cite{onthepower} and \cite{jaeger, jaeger2}) or ii) include all correlations from a specific LCSR computation and accept all hypothesis and assumptions of one specific LCSR computation (followed by \cite{bsz} and \cite{silves}). The observation raised in \cite{onthepower,newpaper} is that option ii) is scheme-independent\footnote{A scheme consists in choosing how to determine the soft FF $\xi_{\perp,\|}$ from combinations of the full FF (see \cite{onthepower}).} at the price of being fully sensitive to all assumptions of a specific LCSR computation embedded into the correlations while {\it option i) is free from LCSR assumptions at the level of the assigned error to these coefficients but at the price of introducing  a scheme dependence}.
  Due to this scheme dependence, an unfortunate choice of scheme can artificially inflate the uncertainties,
as it happens for the observable $P_5^\prime$ in scheme 2 used in ref.\cite{jaeger,jaeger2}, compared to more appropriate schemes like the ones used in \cite{thorsten2} or \cite{onthepower}. 
  This crucial point was missed  in \cite{jaeger,jaeger2} leading them logically to the wrong conclusion   that factorizable power corrections were huge and explained the anomalies observed in this decay.  We illustrate the importance of an appropriate scheme choice not to inflate artificially the error with an example in Table\ref{options}. 
 Needless to say, accepting the possibility of huge power corrections as proposed in \cite{jaeger,jaeger2} would imply a direct conflict with the correlations among FF computed, for instance, in BSZ\cite{bsz}.

\subsection{Hadronic uncertainties for $B \to K^*\mu\mu$: long distance charm contributions}


The common vectorial structure of ${\cal A}_V^\mu$ and ${\cal H}_V^\mu$ is one of the main reasons behind the claim that a large unknown long distance charm contribution maybe the responsible for  {\bf some} of the anomalies. This means that the perturbative SM contribution ${\cal C}^{\rm eff \, SM}_{9\, pert}$ is always associated with a transversity-, process-  and q$^2$-dependent  charm loop contribution $C_9^{\bar c c \, i}(q^2)$
$${\cal C}_{9}^{\rm eff i}(q^2)={\cal C}^{\rm eff\, SM}_{9\,  pert}+ {\cal C}_9^{NP}+ C_9^{\bar c c \, i}(q^2)$$
with $i=\perp,\|,0$. At present only a partial computation by \cite{KMPW} exists based on a 
LCSR computation including the exchange of a single soft gluon. In the notation of \cite{KMPW} the long-distance charm-loop contribution is given by $2 C_1 \tilde g^{\bar c c,B\to K^*, {\cal M}_j}(q^2)$ with $j=1,2,3$ and it is real \cite{private} in the region where it was computed $q^2 \ll 4 m_c^2$. Using a dispersion relation the complete contribution was extended  to the region $1 \leq q^2 \leq 9$ GeV$^2$  (see eq.~7.14 in \cite{KMPW}). In our analysis 
we separate KMPW into the perturbative part and a remaining non-perturbative part. 
We do not use directly the latter, but we consider it as an indication of the magnitude of the effect. Thus, we take
it with  an extra parameter (one per amplitude) allowed to vary from $-1 \leq s_i \leq 1$:    $C_9^{\bar c c \, i}(q^2)=s_i {C_9^{\bar c c \, i}}_{\rm KMPW}$. It is important to remark that our parametrization respects the one used in \cite{KMPW}. 
  \begin{figure} \begin{center}
\includegraphics[width=10cm,height=3cm]{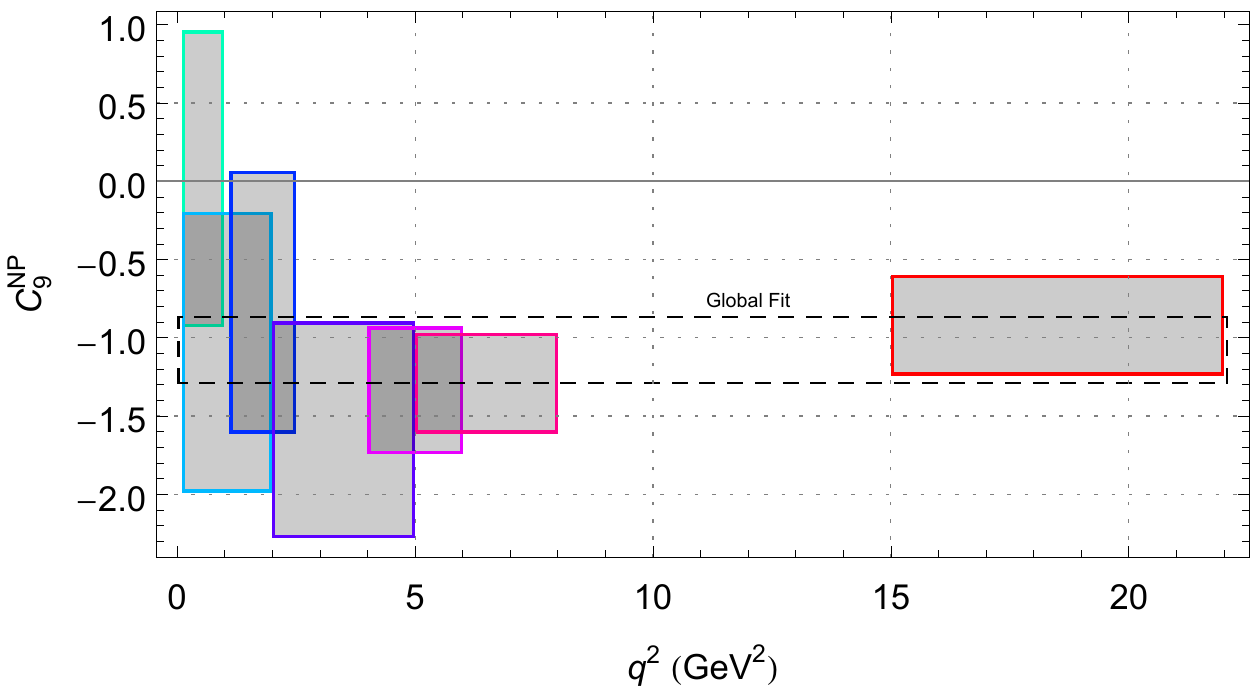}
\caption{Determination of ${\cal C}_9^{NP}$ bin-by-bin from the reference fit. } \label{plotc9bins}
\end{center}\end{figure}
Using this procedure one can estimate, for instance, the charm error associated to the bin [4,6] of $P_5^\prime$ and find that one would need  to artificially enlarge this error by 600\% to get agreement with data \cite{talkconf}. One may also  compare the estimates of the non-factorizable error (including charm) in the long bin ([1,6]) of $P_5^\prime$ for three different papers DHMV'14 ($\pm 0.10$), JC'12 ($\pm 0.09$) and BSZ'15 ($\pm 0.06$). In \cite{global} we performed a test of the implementation of the charm-loop contribution using the global fit. The idea (see also \cite{straub}) is the following: in principle one would expect naturally the charm-loop contribution to be $q^2$-dependent contrary to a universal and $q^2$-independent new physics  ${\cal C}_9^{NP}$. If the charm-loop contribution has been correctly parametrized  and estimated in all modes one should not see any residual $q^2$-dependence in the determination of ${\cal C}_9^{NP}$ bin by bin. Indeed,
 the result of evaluating ${\cal C}_9^{NP}$ using the global fit bin-by-bin \cite{global} does not show any clear signal of a $q^2$ remaining sensitivity. Albeit the errors are still too large to get definite conclusions the comparison of bins [4,6] and [6,8] (where the highest $q^2$-sensitivity was supposed to manifest) are:
$${\cal C}_9^{NP\, [4,6]}=-1.3\pm 0.4 \quad \quad {\cal C}_9^{NP\, [6,8]}=-1.3\pm 0.3 $$
which seems in very good agreement (see Fig. \ref{plotc9bins}). 
The previous test relies strongly on the assumption that NP affects solely ${\cal C}_9^{NP}$. We should also notice that the first bin suffers from lepton-mass effects as discussed in \cite{global}.
We would like to close this section with a comment on a recent paper \cite{silves} advocating for large $c\bar c$ effects. In this paper an arbitrary parametrization is introduced to include a non-factorizable effect: $h_\lambda=h_\lambda^{(0)}+h_{\lambda}^{(1)} q^2 + h_{\lambda}^{(2)} q^4$ (notice the 1 GeV normalization rather than the usual $m_b$) with 18 free parameters. Then a fit to LHCb data is done using only the low-q$^2$ data of $B \to K^*\mu\mu$. First of all, it is not surprising having introduced so many free parameters (including a $q^2$ dependence) that any shape can be fitted making difficult the task to extract any solid conclusion out of that.\footnote{ An example of this difficulty is provided by the authors of \cite{silves}. 
In v1, the authors indicated being able to provide a good description of 
the experimental data. However it turned out that their SM predictions were made on the base of erroneous formulae (with a factor 2 missing in the definition of $S_4$) 
which could be spotted due to an inconsistency among the predictions (see \cite{lathuile}). This mistake in the SM part could be accommodated by the additional polynomial contribution: as this
contribution does not contain physical dynamics, it can easily accommodate any discrepancy between theory and 
an experiment, no matter wheter the discrepancy is real or spurious.} Indeed this paper \cite{silves} contains two different analyses. In one of them, no constraint  is imposed on these parameters and not surprisingly what they found is in good agreement with the results of \cite{global} and \cite{straub}. The result is the presence of a constant universal shift, the same that we called ${\cal C}_9^{NP}$ (that in our case also explains $R_K$ 
free from long-distance charm contributions in the SM) while in their interpretation they attribute it  to a completely unknown lepton-flavour and $q^2$-independent  non-factorizable effect (missing of course the possibility to explain $R_K$ and any future LFUV observables). However, on the second part of \cite{silves} they impose  the constrain that the parameters should fulfill the KMPW charm-loop partial computation for $q^2 \leq 1$ GeV$^2$  effectively imposing the SM value for the prediction at low $q^2$. This generates several problems, first it tilts artificially the fit, second  the constrain is imposed in the region where lepton-mass effects maybe relevant and finally, the KMPW charm-loop contribution is real\cite{private} while the one in \cite{silves} is complex in this region. More details can be found in \cite{newpaper}. On top of all these conceptual problems their computation of the $S_i$ observables  was done using BSZ form factors with an out-of-date input\footnote{
The authors of Ref.~\cite{bsz} spotted a problem in one of the inputs used to evaluate BSZ FF coming from the literature and that affected twist-4 ${\cal O}(\alpha_s)$ terms. This was corrected and the $S_i$ were reevaluated in the last version of Ref \cite{bsz}.}
 and would require a reevaluation (even if this is not the most relevant issue).

\section{What's next? LFUV observables $Q_i$}
One natural question to ask is why the pull-SM of a NP contribution to the coefficient of the semileptonic operator $\mathcal{O}_9^{}=\frac{\alpha}{4\pi}[\bar{s}\gamma^\mu P_{L}b]
   [\bar{\mu}\gamma_\mu\mu]$ is so large appearing as the main responsible of all anomalies and tensions. The answer is easy to understand by looking at Table 3. While ${\cal C}_9^{NP \, \mu}$ consistently reduces the tension in all anomalies shown in Table 3 the other Wilson coefficients go in different directions for each tension requiring cancellations to improve the data. A consequence of this is that the days of ``barring accidental cancellations" are gone.

 \begin{table}  \begin{center}{\small \begin{tabular}{ccccccccc}    \label{c9alwayspp} 
  &  & $ R_K$ &  $\langle P_5^\prime\rangle_{[4,6], [6,8]} $ &  ${\cal B}_{B_s \to \phi\mu\mu}$  &  ${\cal B}_{B_s \to \mu\mu}$ \small  & low-recoil & best-fit-point  \\\hline\hline
\multirow{2}{*}{$\C9^{NP}$} &$+$& & &  & & \\
&$-$ & $\checkmark$ & $\checkmark$ & $\checkmark$  & & $\checkmark$ & { X} \\ \hline
\multirow{2}{*}{$\C{10}^{NP}$} &$+$& $\checkmark$&  & $\checkmark$   & $\checkmark$  & $\checkmark$ & { X}  \\
&$-$ & & $\checkmark$  &  &  & \\ \hline
\multirow{2}{*}{$\C{9^\prime}$} &$+$&   &   & $\checkmark$ & & $\checkmark$  & { X} \\
&$-$ & $\checkmark$ & $\checkmark$ & & &\\ \hline
\multirow{2}{*}{$\C{10^\prime}$} &$+$& $\checkmark$ & $\checkmark$ &  & &  \\
&$-$ &  &   & $\checkmark$& $\checkmark$ & $\checkmark$ & { }  \\ \hline
& & & &  But also ${\cal C}_7^{NP}, {\cal C}_7^\prime, ....$   &   && \\ \hline
\end{tabular}}
\end{center} 
\caption{Anomalies versus variation of one single Wilson Coefficient. A check-mark indicates that a shift in the Wilson coefficient with this sign moves the prediction in the right direction to alleviate the tension.}
\end{table}

 One might be tempted after a fast (and superficial) analysis to arrive to the conclusion that long distance charm is mimicking a fake NP contribution  and that we basically have to accept that our understanding of charm physics at present is near null, since a first estimate of its contribution (as discussed in the previous section) falls short by 600\%. Basically the question would be between NP {\bf or} an unknown huge charm contribution. However, this conclusion is invalidated if one takes a more global (and deep) view of {\bf all} the experimental data. Under the hypothesis that there is LFUV as pointed preliminary by $R_K$ then the long distance charm explanation is seriously in trouble. LFUV observables by construction in the SM are insensitive to long-distance charm \cite{setlfv}. In other words, if one explains $R_K$ via a ${\cal C}_9^{NP \, \mu}$ contribution one is able also to explain $P_5^{\prime\mu}$ as well as other tensions (see Fig.\ref{impact}), leaving the space for an unknown long-distance charm contribution quite reduced. Following this line of reasoning we can naturally construct a set of LFUV observables, that we called $Q_i=P_i^\mu-P_i^e$~\cite{Capdevila:2016ivx} that inherits the interesting properties of optimized observables and 
 have little sensitivity to hadronic uncertainties in the SM  (in particular long-distance charm). A clear deviation from zero in those observables is an unquestionable signal of  NP violating lepton-flavour universality. Obviously, in presence of NP the problem of hadronic uncertainties reemerges, but 
 then we enter a completely different field, that of NP Discovery. 
  At this point  the previous question on NP or charm turns into  NP {\bf times} long distance charm ($P_i^\mu-P_i^e=(C_i^\mu-C_i^e )\times {\rm hadr.}$) and the problem is not anymore the discovery of NP but disentangling the correct scenario. It is precisely for this reason why it is essential to use optimized observables. We can illustrate this idea with a simple example, the comparison between $P_5^{\prime\mu}$ and $Q_5$.
Both observables are  independent at LO on soft FF, but while $P_5^{\prime\mu}$ includes a long-distance charm contribution as described in the previous section, $Q_5$ is totally insensitive to it in the SM. It is interesting to compare  the error size of the SM predictions of both observables in the bin [4,6], while $\langle P_5^{\prime \mu}\rangle_{[4,6]}=-0.82\pm 0.08$, the LFUV associated observable\footnote{The hat means that this prediction takes into account the  way LHCb measures $F_L$ (see \cite{Capdevila:2016ivx} for definitions). 
 If this effect is not considered the error size would be below 10$^{-3}$! See the error of the $Q_5$ observable (with no hat) in \cite{Capdevila:2016ivx}.}
 $\langle\hat{Q_5}\rangle_{[4,6]}\!=\!-0.002\pm 0.017\!$ has an error smaller by near a factor of 5. But also in presence of NP in ${\cal C}_9^{NP\mu}$ the error of $\hat{Q}_5$ is less than a half the one of $P_5^\prime$.

\begin{figure}
\hspace*{-0.4cm}\includegraphics[width=5cm,height=4.2cm]{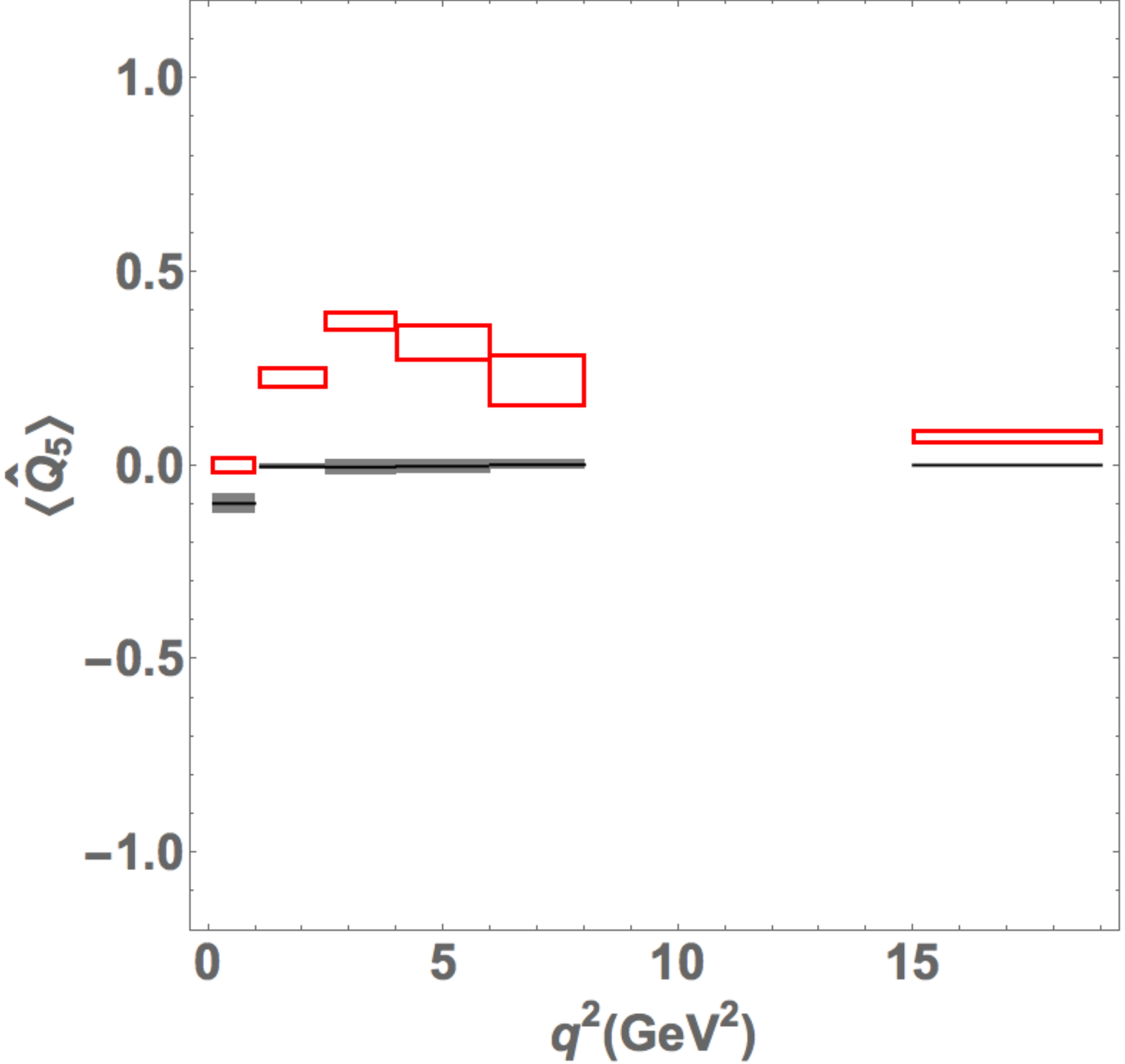}\hspace*{-0.7cm}
  \includegraphics[width=4.1cm,height=4.7cm]{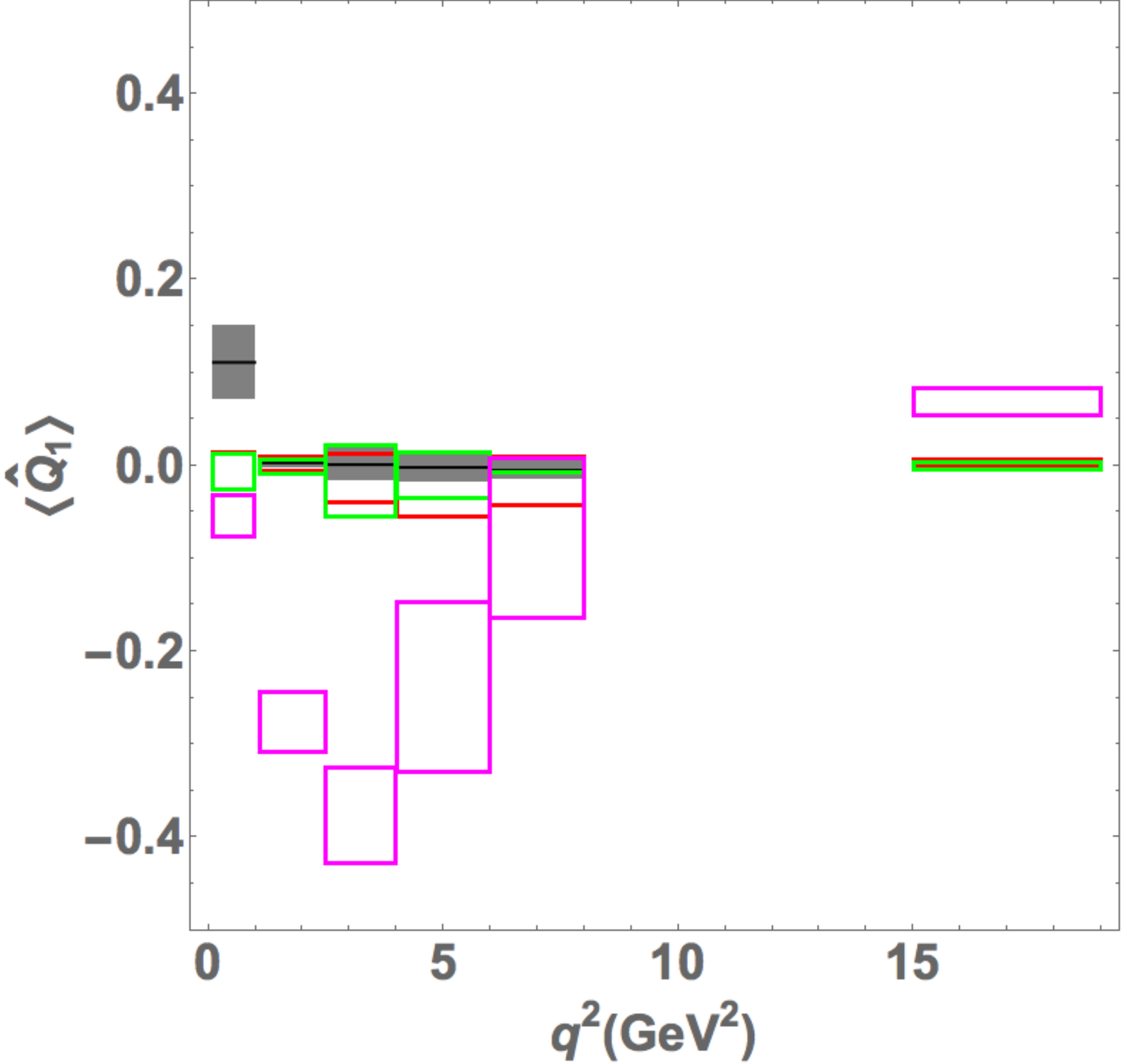}\hspace{-0.7cm}
    \includegraphics[width=4.1cm,height=4.7cm]{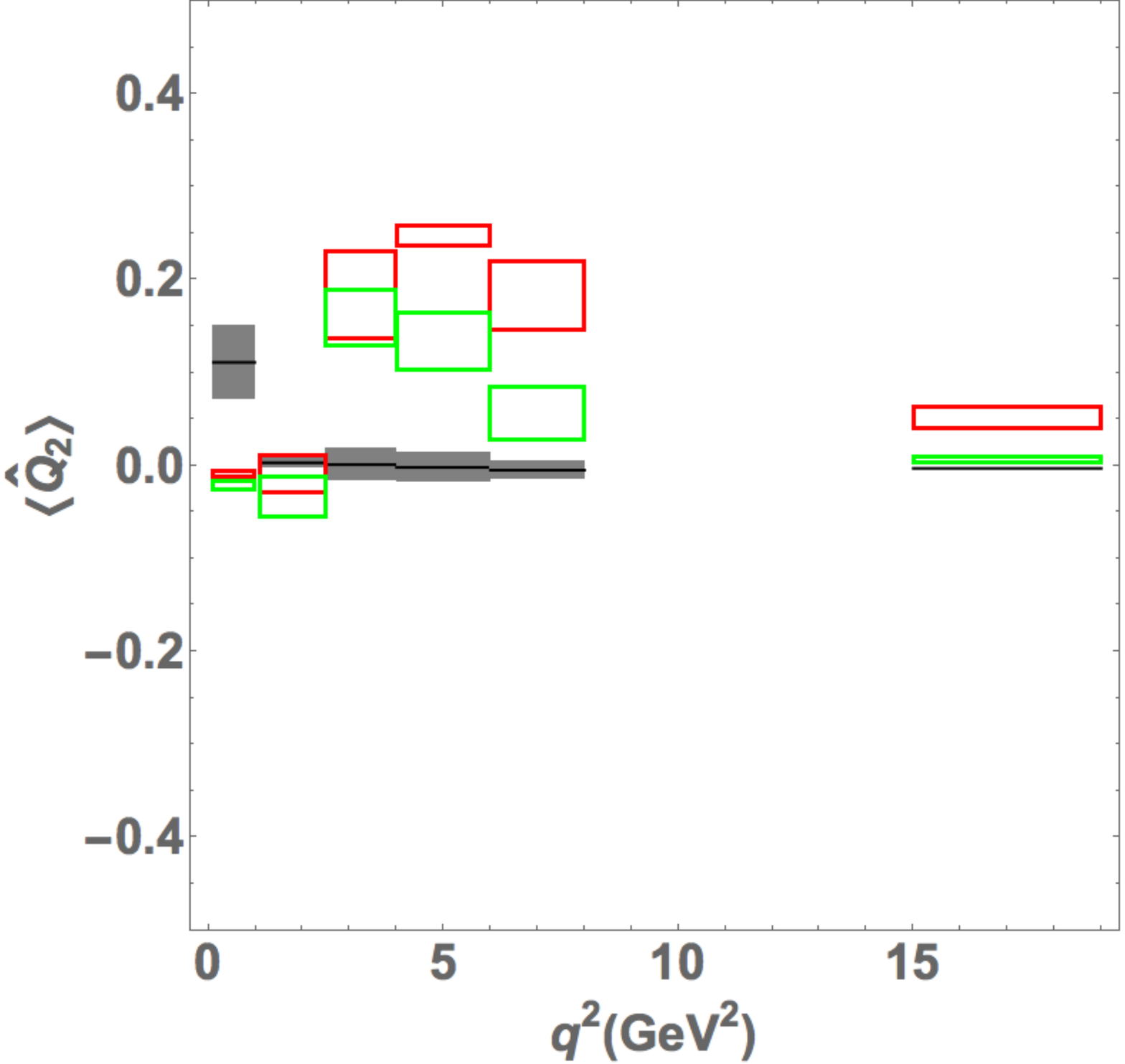}\hspace{-0.7cm}
    \includegraphics[width=4.1cm,height=4.7cm]{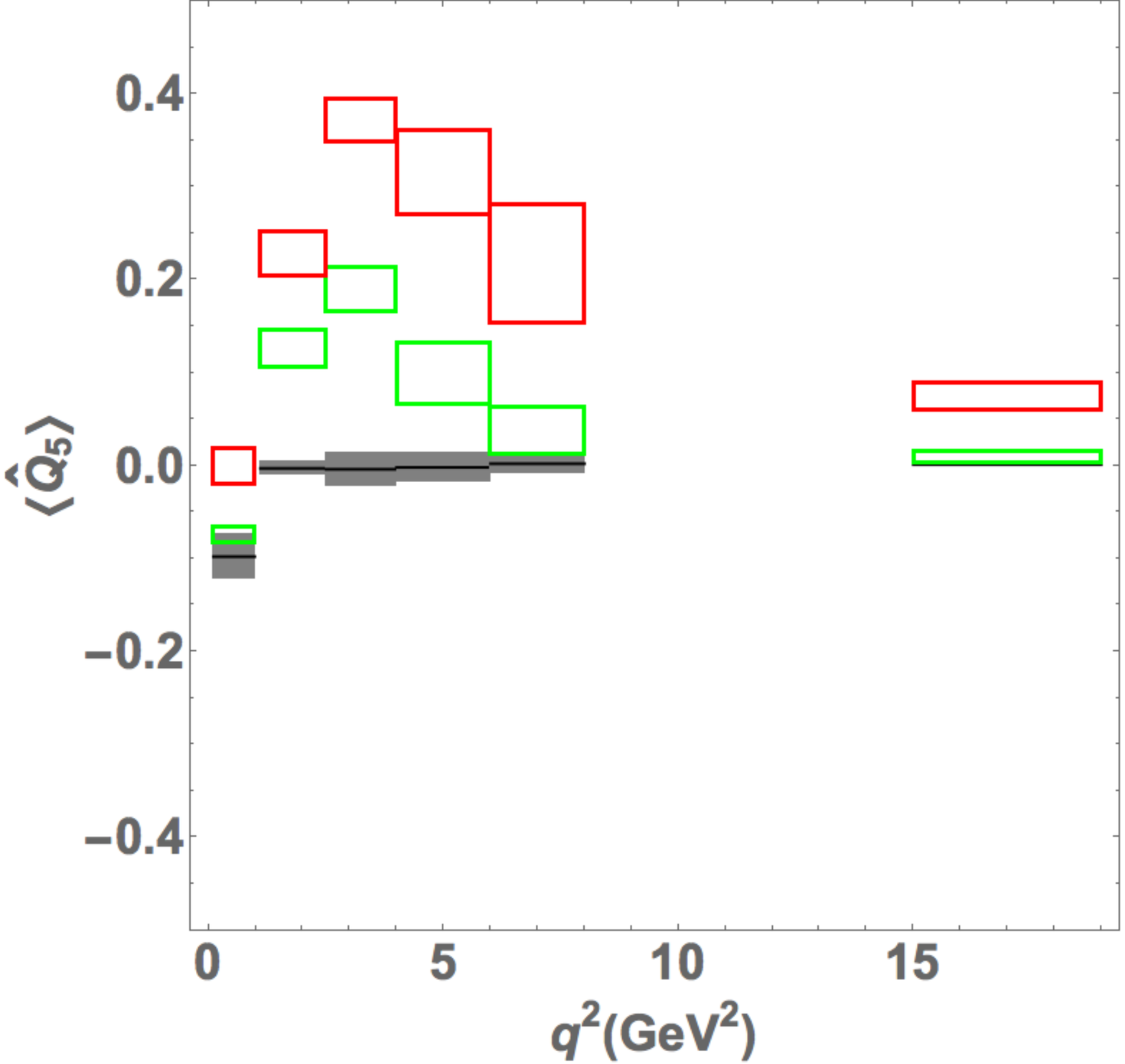}   
   \caption{(1st plot) $\hat{Q}_5$ to compare with $P_5^\prime$ of Fig. 3 (same conventions and scale).
    (2nd-4th plot) Example of the impact of scenario 1 (in red: ${\cal C}_9^{NP\mu}<0$) and 2 (in green: ${\cal C}_9^{NP\mu}=-{\cal C}_{10}^{NP\mu}$) for observables 
    $\hat{Q}_{1,2,5}$ and scenario 4 only for $\hat{Q}_1$ (in magenta for NP also in $\C{9'}^{NP\mu}$, 
    $\C{10'}^{NP\mu}$). See \cite{Capdevila:2016ivx} for details on scenarios.     
    }\label{qi}
\end{figure}

The angular analysis of a subset of $Q_i$ observables ($i=1,2,4,5$), as shown in Fig.\ref{qi} (only some of them are displayed, see \cite{Capdevila:2016ivx} for more details), may be enough to distinguish a right handed current signal at low-$q^2$ in $Q_1$ and $Q_4$ or get a hint of the scenario that goes beyond ${\cal C}_9^{NP}$, focusing at the very low and very high $q^2$ values inside the large-recoil region of the observables $Q_2$ and $Q_5$. 

In summary, in the near future, the expected CMS (and ATLAS)  measurement of $P_5^\prime$, 
together with an  update of the measurement of the $P_i$ observables and $R_K$ with higher integrated luminosity,
and a first measurement of $R_{K^*}$ and $R_\phi$  will be relevant steps to push  the discrepancy of $b\to s\ell\ell$ modes w.r.t. SM beyond 5$\sigma$.
 However, the first precise measurement of the subset of $Q_i$ observables $i=1,2,4,5$ may cause a major breakthrough in our search for NP, helping to disentangle the final scenario of NP beyond ${\cal C}_9^{NP\mu}$. In this search LHCb will share a leading role  with Belle-II.

\section{Conclusions}
Exciting times are ahead of us in the near future. Belle-II, LHCb, CMS and ATLAS will provide new insights on the flavour anomalies soon. At present LHCb data on $b \to s\ell\ell$ decays shows tensions with SM in several observables: $P_5^\prime$, $BR(B_s \to\phi\mu^+\mu^-)$, $R_K$ and several low-recoil bins. Global fits show a clear preference at present for a scenario with a large negative ${\cal C}_9^{\rm NP\mu}$ contribution with a discrepancy w.r.t. SM above 4$\sigma$.  A bin-by-bin analysis  shows a clear compatibility of the fit with a $q^2$-independent effect, albeit the errors are still large. We presented also an analysis of some alternative solutions proposed in the literature to solve a few of these discrepancies within the SM.

Finally, we discussed observables sensitive to the violation of lepton flavour universality and free from hadronic uncertainties in the SM, like $R_K$ or the recently proposed $Q_i$ (more observables are defined in \cite{Capdevila:2016ivx}). These new observables 
 represent the next step in disentangling the final scenario of LFUV, where other Wilson coefficients besides ${\cal C}_9^{\rm NP\mu}$ will be naturally switched on.
\medskip

{\it Acknowledgments:}
This work received financial support from the grant FPA2014-61478-EXP [SDG, JM, JV];
 from the grants FPA2013-46570-C2-1-P and 2014-SGR-104, and from the Spanish MINECO under the project MDM-2014-0369 of ICCUB (Unidad de Excelencia ``Mar\'ia de Maeztu'') [LH]; from
 DFG within research unit FOR 1873 (QFET) and from CNRS [JV]; from the EU Horizon 2020 programme from the grants No 690575, No 674896 and No. 692194 [SDG].


\begin{thebibliography}{99}

\bibitem{rk}
  {\bf LHCb} Collaboration,
  Phys.\ Rev.\ Lett.\  {\bf 113} (2014) 151601.

\bibitem{babarrk}
  J.~P.~Lees {\it et al.} [BaBar Collaboration],
  Phys.\ Rev.\ D {\bf 86} (2012) 032012


\bibitem{belle}
  M.~Huschle {\it et al.} [Belle Collaboration],
  Phys.\ Rev.\ D {\bf 92} (2015) no.7,  072014

\bibitem{lhcb}
  R.~Aaij {\it et al.} [LHCb Collaboration],
  Phys.\ Rev.\ Lett.\  {\bf 115} (2015) no.11,  111803
   Addendum: [Phys.\ Rev.\ Lett.\  {\bf 115} (2015) no.15,  159901]
  [arXiv:1506.08614 [hep-ex]].

\bibitem{p5p}
  S.~Descotes-Genon, J.~Matias, M.~Ramon and J.~Virto,
  JHEP {\bf 1301} (2013) 048

 \bibitem{opt1}
  F.~Kruger and J.~Matias,
  Phys.\ Rev.\ D {\bf 71} (2005) 094009.
  
  \bibitem{opt2}
  J.~Matias {\it et al.}
  JHEP {\bf 1204} (2012) 104.
    
\bibitem{opt3}
  S.~Descotes-Genon {\it et al.},
  JHEP {\bf 1305} (2013) 137.

\bibitem{2013lhcb}
  {\bf LHCb} Collaboration,
  PRL\  {\bf 111} (2013) 191801.


\bibitem{2015lhcb}
  R.~Aaij {\it et al.} [LHCb Collaboration],
  JHEP {\bf 1602} (2016) 104.

\bibitem{belle2016}
  A.~Abdesselam {\it et al.} [Belle Collaboration],
  arXiv:1604.04042 [hep-ex].
  
  
  
\bibitem{lhcbphi}
  R.~Aaij {\it et al.} [LHCb Collaboration],
  JHEP {\bf 1509} (2015) 179
  [arXiv:1506.08777 [hep-ex]].
    
  
\bibitem{bsz}
  A.~Bharucha, D.~M.~Straub and R.~Zwicky,
  arXiv:1503.05534 [hep-ph].


\bibitem{global}
  S.~Descotes-Genon {\it et al},
  JHEP {\bf 1606} (2016) 092.

  
  \bibitem{bulk}
  A.~Ali, G.~F.~Giudice and T.~Mannel,
  Z.\ Phys.\ C {\bf 67} (1995) 417.

  G.~Hiller and F.~Kruger,
  Phys.\ Rev.\ D {\bf 69} (2004) 074020.

    S.~Descotes-Genon {\it et al.} ,
  JHEP {\bf 1106} (2011) 099.

    W.~Altmannshofer, P.~Paradisi and D.~M.~Straub,
  JHEP {\bf 1204} (2012) 008.

    C.~Bobeth, G.~Hiller and D.~van Dyk,
  Phys.\ Rev.\ D {\bf 87} (2013) 034016.



\bibitem{anomaly}
  S.~Descotes-Genon, J.~Matias and J.~Virto,
  Phys.\ Rev.\ D {\bf 88} (2013) 074002.


\bibitem{Altmannshofer:2013foa}
  W.~Altmannshofer and D.~M.~Straub,
  Eur.\ Phys.\ J.\ C {\bf 73} (2013) 2646.
 
\bibitem{Beaujean:2013soa}
  F.~Beaujean, C.~Bobeth and D.~van Dyk,
  EPJC {\bf 74} (2014) 2897
   [EPJC {\bf 74} (2014) 3179].



\bibitem{straub}
  W.~Altmannshofer and D.~M.~Straub,
  Eur.\ Phys.\ J.\ C {\bf 75} (2015) 8,  382;
  arXiv:1503.06199 [hep-ph].


\bibitem{Hurth} 
T.~Hurth, F.~Mahmoudi and S.~Neshatpour,
  Nucl.\ Phys.\ B {\bf 909} (2016) 737.

\bibitem{thorsten}
  M.~Beneke, T.~Feldmann and D.~Seidel,
  Nucl.\ Phys.\ B {\bf 612} (2001) 25,


\bibitem{thorsten2}
  M.~Beneke, T.~Feldmann and D.~Seidel,
  EPJC {\bf 41} (2005) 173


\bibitem{onthepower}
  S.~Descotes-Genon  {\it et al.},
  JHEP {\bf 1412} (2014) 125.

\bibitem{KMPW}
  A.~Khodjamirian  {\it et al.},
  JHEP {\bf 1009} (2010) 089.



\bibitem{jaeger}
  S.~J\"ager and J.~Martin Camalich,
  JHEP {\bf 1305} (2013) 043.
  
  
 \bibitem{newpaper}  
B.~Capdevila, S. Descotes-Genon, L. Hofer and J. Matias in preparation.


\bibitem{charles}
  J.~Charles, A.~Le Yaouanc, L.~Oliver, O.~Pene and J.~C.~Raynal,
  Phys.\ Rev.\ D {\bf 60} (1999) 014001




\bibitem{silves}
M.~Ciuchini {\it et al.},
  JHEP {\bf 1606} (2016) 116.



\bibitem{jaeger2}
  S.~J\"ager and J.~Martin Camalich,
  Phys.\ Rev.\ D {\bf 93} (2016) no.1,  014028.

\bibitem{private}
A. Khodjamirian private communication.


\bibitem{talkconf}
J. Matias, Talk at Fourth Annual Large hadron Collider Physics, Lund, June 2016.

\bibitem{lathuile}
J. Matias, Talk at Les Rencontres de Physique de la Vallee d'Aoeste, La Thuile, March 2016.




  
  
  
  

\bibitem{setlfv}
  G.~Hiller and M.~Schmaltz,
  Phys.\ Rev.\ D {\bf 90} (2014) 054014;
  JHEP {\bf 1502} (2015) 055.

  D.~Ghosh, M.~Nardecchia and S.~A.~Renner,
  JHEP {\bf 1412} (2014) 131.

  T.~Hurth, F.~Mahmoudi and S.~Neshatpour,
  JHEP {\bf 1412} (2014) 053.



  D.~Be\v{c}irevi\'c, S.~Fajfer and N.~Ko\v{s}nik,
  Phys.\ Rev.\ D {\bf 92} (2015) no.1,  014016.


\bibitem{Capdevila:2016ivx}
  B.~Capdevila {\it et al.},
  arXiv:1605.03156 [hep-ph].
  
  
  
  
\end{thebibliography}
\end{document}